# Glassy Li Metal Anode for High-Performance Rechargeable Li Batteries

Xuefeng Wang[1]†, Gorakh Pawar[2]†, Yejing Li[1], Xiaodi Ren[3], Minghao Zhang[1], Bingyu Lu[1],

Abhik Banerjee[1], Ping Liu[1], Eric J. Dufek[4], Ji-Guang Zhang[3], Jie Xiao[3], Jun Liu[3], Ying Shirley

Meng[1]*, Boryann Liaw[4]*

[1] Department of NanoEngineering, University of California San Diego, 9500 Gilman Drive, La

Jolla, CA 92093, USA

[2] Department of Material Science and Engineering, Idaho National Laboratory, 1955 N. Fremont

Avenue, Idaho Falls, ID 83415, USA

[3] Energy and Environmental Directorate, Pacific Northwest National Laboratory, 902 Battelle

Boulevard, Richland, Washington 99354, United States

[4] Department of Energy Storage and Advanced Vehicles, Idaho National Laboratory, 1955 N.

Fremont Avenue, Idaho Falls, ID 83415, USA

* Correspondence to: boryann.liaw@inl.gov (B. Liaw); shirleymeng@ucsd.edu (Y.S. Meng)

† GP and XW are co-first authors, who made equal contributions to the major content of the

article.

**Abstract:**

Controlling nanostructure from molecular, crystal lattice to the electrode level remains as arts in practice, where nucleation and growth of the crystals still need more fundamental understanding and precise control to shape the microstructure of materials and their properties. It is vital to achieve dendrite-free Li metal anodes with high electrochemical reversibility for practical high-energy rechargeable Li batteries. Here, cryogenic-transmission electron microscopy was used to reveal the evolving nanostructure of Li metal deposits at various transient states in the nucleation and growth process, in which a disorder-order phase transition (DOPT) was observed as a function of current density and deposition time. The atomic interaction over wide spatial and temporal scales was depicted by the reactive-molecular dynamics simulations to assist the understanding of the kinetics. Compared to crystalline Li, glassy Li outperforms in electrochemical reversibility and it is a desired structure for high-energy rechargeable Li batteries.

**Main Text:**

Metal-based anodes, such as lithium (Li) metal can lead to the highest specific energy for rechargeable batteries and show great promise to meet the increasing demand for energy storage applications in the future. However, most Li metal electrodes suffer from poor electrochemical reversibility and short cycle life. Since the properties of the Li metals (e.g. nano-/microstructure, morphology and electrochemical performance) in the rechargeable batteries are largely governed by the electrochemical process, it is critical to comprehend the underlying mechanism of Li deposition by both experimental and theoretical work, especially in the very early stage of nucleation to seek better control of the Li deposition.

To explore the electrodeposition behavior of Li metal, various experimental techniques have been developed to capture the kinetic process and probe the structural evolution of Li metal at different stages and conditions, as displayed in Fig. S1 and Table S1. Although *in situ*, *operando* microscopies allow to visualize the real-time microstructure evolution [1, 2, 3, 4, 5, 6, 7], it is still very difficult to capture the initial stage of Li deposition when Li metal begins to nucleate and subsequently grow into stable microstructures. In order to push the detection limit to nano or even atomic scale, cryogenic protection is essential to minimize the beam damage, while preserving the intrinsic structure of Li deposits. Recently, the cryogenic-electron microscopy (cryo-EM) has been proved useful to study the nanostructures of Li metal and reveal the variations in its crystallinity and the solid electrolyte interphases (SEI) [8, 9]. Such variations may significantly alter the growth of Li deposits and their physicochemical properties, yet detailed understanding has not been established before.

Classical nucleation theory (CNT) depicts that nuclei would appear if the embryo's bulk energy overcomes the surface energy and the nuclei would grow if the size of embryo exceeds the critical radius [10]. For electrochemical nucleation, this process is driven by the charge transfer and sustained by the mass transport of ions near the electrolyte-electrode interface [11]. Previous models based on these principles [12, 13, 14, 15, 16, 7] underline some key concepts like free energy, surface tension and overpotential, which are macroscopic properties; but these concepts become ambiguous when a nucleus only comprises a few atoms. These properties also do not describe the atomic interactions, attachment and detachment kinetics to small clusters with sufficient details to delineate different nanostructure configurations, which is urgently required to develop microscopic and even atomistic models to understand the Li nucleation and growth.

Here, we applied cryo-EM to capture the kinetic progression of nucleation of Li metal and use reactive-molecular dynamics (*r*-MD) simulations to understand the atomic interaction inside. A disorder-order phase transition (DOPT) was revealed and explained as a function of current density and deposition time. On this basis, the crystallinity of nuclei was correlated with the subsequent growth of the nanostructure and morphology that are pertinent to the electrochemical performance of the Li metal electrode.

**Cryo-TEM Observations**

The cryo-TEM results (Fig. 1 and Fig. S2 [17]–Fig. S12) show the nano- and microstructure evolution of electrochemically deposited Li (EDLi) as a function of deposition time (5–20 min, Fig. S3, S4 and S6) and current density (0.1–2.5 mA cm[−2], Fig. S5 and S10). At 0.5 mA cm[−2], although the deposition time was as short as 5 min, tubular EDLi was formed about 200 nm in diameter and >1 µm in length (Fig. S4). Interestingly, no lattice fringes (Fig. 1a and Fig. S6a-d), nor characteristic bright diffraction spots/rings of the body-centered cubic (*bcc*) Li metal packing in the Fast Fourier transformed (FFT) image (Fig. 1d) were found, suggesting that the EDLi is amorphous or glassy. The local structure in Fig. 1a displays the disordered arrangement of Li atoms (comparison to those in Fig. 1c and Fig. S9). The metallic nature of EDLi is confirmed by the electron energy loss spectrum (EELS) (Fig. S7) and electrochemical reversibility (Fig. S8) [18]. At 10 min, a small portion of EDLi near the surface became crystalline (Fig. 1b and Fig. S6e-h) as evident by the characteristic bright diffraction spots (highlighted by the red arrows) in Fig. 1e with weak contrast. Lattice fringes were present in this region (Fig. 1b) and the distance between the adjacent fringes was measured to be 0.246 nm, consistent with the spacing between Li (110) planes. It is worth noting that the domain of the crystalline EDLi (shown by the white dash line) is about 5 nm in size, a size the $r$-MD simulation predicted the transition to the crystalline phase is eminent and spontaneous. The crystallinity degree of the EDLi is further enhanced with increased deposition time. For 20 min, large crystalline domains (>50 nm) and intense characteristic bright diffraction spots of the $bcc$ structure were present (Fig. 1c, f, and Fig. S6i-l). Thus, at the same deposition rate, as the deposition time increases more Li atoms aggregate to form larger embryos. Once an embryo is larger than a critical size, the DOPT occurs.

Further studies of the nanostructure dependence on current density (Fig. 1g-l for constant capacity and Fig. S10), we found that the higher the current density was used, the larger the embryo size and the higher the degree of crystallinity in EDLi (the statistical analysis is shown in Fig. S10). The EDLi is mostly amorphous at 0.1 mA cm$^{-2}$ (Fig. 1g, j, and Fig. S10a, e, i, and m), partly crystalline at 1.0 mA cm$^{-2}$ (Fig. 1h, k, and Fig. S10c, g, k, and o) and highly crystalline at 2.5 mA cm$^{-2}$ (Fig. 1i, l, and Fig. S10d, h, l, and p). The crystalline domains of EDLi at 1.0 mA cm$^{-2}$ were found near the surface (labeled by the red square in Fig. S10c) and about 5 nm in size (labeled by the white dash line in Fig. 1h). These results suggest that: (1) the embryo size is key to the DOPT, (2) matured crystallites are often larger than 5 nm, and (3) the embryo size is sensitive to the current density. Thus, the high current densities expedite the Li aggregation to precipitate the crystalline nuclei. This result implies the diverse local current density may lead to a wider distribution of Li embryo sizes with different degrees of crystallinity and domain sizes (Fig. S5).

Obtaining glassy EDLi during electrodeposition is quite surprising for alkali metals but not preposterous, since DOPT was also observed in other metals (such as Ni, Au, Ag, Pt$_3$Co, FePt, etc.) during nucleation and growth in solid or solution with precursors via $in$ $situ$ annealing and by liquid-cell electron microscopy [19, 20, 21, 22, 23]. These DOPTs deviate from the single-step process explained by the CNT and suggest the multi-step nature of crystallization through amorphous interphases. It is worth emphasizing that all the DOPT critical size is around 2-5 nm.

**Reactive Molecular Dynamics Simulations**

To understand this peculiar amorphous nature of the EDLi, a three-stage *r*-MD protocol was used to simulate the nucleation process with a variety of discrete canonical ensembles and conditions typically comprising (1) heating at 500 K for 0.1 ns to provide sufficient driving force for the nucleation, (2) quenching to 300 K with a cooling rate of 1 K ps–1, and (3) equilibrating at 300 K for 5 ns (Fig. 2a). The reactive force field ReaxFF potential [24] was used to simulate the Li-Li interactions and LAMMPS package [25, 26] to perform all simulations (Fig. S13-17) [27, 28, 29, 30, 31]. It is worth mentioning that in the heating and quenching stage, the nucleation sites dynamically evolved into more stable embryos, as Ostwald ripening explains (Fig. 2a) [32]. Such a condensation process requires a specific "incubation time" to give the birth of a stable embryo, which is overlooked by the quasi-equilibrium models including the CNT or density function theory to date. Intriguingly, all embryos studied here were disordered at the end of quenching, far away from the stable, ordered crystalline state. Thus, the incubation and condensation process is key to dictating the time to reach a stable lattice structure, depending on the ensemble size and the initial energy state given (Fig. 2b-d and Fig. S13).

Fig. 2b illustrates a sampling of the lattice structure of Li embryos as a function of embryo size at the end of the simulation. Lithium embryos with >700 Li atoms in the ensemble (at a packing density of $\rho = 0.0534$ g cm–3) are able to transform into a nucleus with the *bcc* lattice structure, whereas those <700 Li atoms remain disorder/amorphous. Embryos of 700 Li atoms exhibit either amorphous or crystalline states with a broad range of kinetic pathways and incubation time (Fig. 2c). Thus, embryo of 700 Li atoms, a size about 2-3 nm, sets the threshold for the DOPT. Increasing the current density yields larger embryo size, which rapidly reduce the incubation time for DOPT (Fig. 2c) and increase the fraction of *bcc* ordered/crystalline Li in the lattice (Fig. 2d). If the current density is lower than a threshold, the incubation time for the DOPT is on the order of *ns*, which is rarely emphasized by the CNT [33, 34].

As Li aggregates, the incubation and physical spatial confinement from the neighboring bodies shall determine if an embryo has sufficient exergy (primarily entropy), mobility and time to carry out the DOPT. Fig. S14 illustrates the variability in the kinetic pathways and temporal nanostructure evolutions in terms of packing density, mass and energy exchange. At a low packing density of 0.0534 g cm$^{-3}$ (Fig. S14a), whether mass or energy exchange was confined or not, the Li aggregation produces mostly amorphous phases. As the current density increases, the packing density of Li atoms shall increase in proportion (Fig. S14b-d). Further confinement in mass and energy exchange shall expedite the DOPT and shorten the incubation time. Diverse kinetic results shall determine the microstructure and morphological evolutions: from spherical amorphous nanostructures of a random distribution of sizes to a host of microstructures of diverse crystallinity, including sheets and rods that comprise amorphous nanostructures (Fig. S14a and amorphous part of Fig. S14b), mixtures of crystalline nanostructures of various sizes and shapes (crystalline part of Fig. S14b and c), and connected networks of micro-grains and micro-pores (Fig. S14d). Such a dynamic range of atomic lattice arrangements further affects the subsequent larger size microstructure evolutions, crystallinity, and morphological and shape changes in a solid particle. A larger scale (in the 100 nm) of simulation and representation of this behavior is shown in Fig. S17.

**From Nucleation to Growth**

The microstructure and morphology of the Li deposited to 1 mAh cm$_{-2}$ loading were examined by SEM and cryo-TEM. The results in Fig. 3 are consistent with the simulation predication. Most of the Li metals grown at 0.1 mA cm$_{-2}$ have sheet-like morphology and their planar size can be as large as several micrometers (Fig. 3a, d and g). This large sheet is originated from the amorphous nature of the Li nuclei, which mostly remained at the end of the growth (Fig. 3j). In contrast, the Li metals grown at 0.5 and 2.5 mA cm$_{-2}$ contain about 30% (Fig. 3k) and 80% (Fig. 3l) crystalline fragments respectively, of which the distribution is highlighted in the gold color. The crystalline Li (c-Li) fragment at 0.5 mA cm$_{-2}$ is sandwiched between the surface SEI and the amorphous bulk Li (a-Li) (Fig. 3k), while the c-Li dominates the Li deposits at 2.5 mA cm$_{-2}$ (Fig. 3l). As a result, ribbon- and even dendrite-like Li deposits are formed at 0.5 (Fig. 3b, e and h) and 2.5 mA cm$_{-2}$ (Fig. 3c, f and i) with rapidly reduced planar diameter and lower Coulombic efficiency (Fig. S8). Thus, it is reasonable to correlate the microstructure, morphology and performance of Li deposits with the order-disorder nanostructure of the Li nuclei: the higher the crystallinity, the finer the final shape, the lower the electrochemical reversibility. The initial nucleation shapes the subsequent growth of the Li metal deposits.

**Strategies and Applications**

Fig. 4. illustrates the benefits of glassy Li metal as a rechargeable Li battery anode. The glassy nature (absence of the ordered nanostructure, grain boundaries and crystal defects) of Li metal avoids the epitaxial growth and enables the multi-dimensional growth into large grains, which is the desired form for a practical Li metal anode. The Li grains have the higher density, lower porosity and tortuosity, less reactivity and better microstructure interconnections than the dendrites. These features can significantly minimize the volume expansion, reduce the side reaction between Li and electrolyte, and maintain an effective electronic and ionic network or percolation pathway. As a consequence, a higher electrochemical reversibility would be expected for glassy Li. Therefore, from a structural perspective, glassy Li metal may be the key to solve the long-standing cyclability issue of Li metal electrode for high-energy rechargeable Li batteries.

As demonstrated above, the DOPT is regulated by the packing density, energy and mass transfer during nucleation process, which in turn provide the ways to facilitate the formation of favorable glassy Li metal. Reducing the current density down to the critical point will directly decrease the initial packing density and facilitate a longer incubation. An alternative way is to use the three-dimensional (3D) substrate (current collector) to reduce the effective local current density. Meanwhile, fast ion and electron conduction is advantageous to reduce mass and energy transfer barriers during nucleation and growth process, which requires electrolyte design and interphase engineering. These strategies have been proved effective to obtain larger grains of Li with enhanced reversibility, although prior arts have not been able to correlate the intrinsic nature of amorphous or glassy Li with the improved performance. Herein, we provide a microscopic perspective on the working principle of these strategies and propose that they could actually alter the nano- and microstructure of the Li deposits.

As a proof of concept, Fig. 5 shows that the electrolyte design indeed have dramatic influence on the bulk lattice structure (Fig. 5c-e) of the Li deposits, which regulates the final morphology (Fig. 5a and b) and electrochemical performance (Fig. 5f). Statistical analysis results (Fig. 5c) indicate that the advanced electrolyte is in favor of forming amorphous Li deposits (~76.8% a-Li *vs.* 23.2% c-Li), while the crystalline ones are dominant in the baseline electrolyte (~48.4% a-Li *vs.* 52.6% a-Li). For their nanostructure, the lattice structure is predominant disorder in the former (Fig. 5e), but a mixture of order and disorder segments appeared in the latter (Fig. 5d); whereas both display similar surface SEI structure (~10 nm thick and composed of $Li_2O$ nanocrystals; black regime). This result demonstrates that it is the intrinsic nanostructure property of Li that governed the final shape and performance of the Li metal anode. As a result, long and thin ribbon-like Li deposits (Fig. 5a) were formed in the baseline electrolyte while large Li chunky deposits found in the advanced electrolyte (Fig. 5b). In Li-free Cu‖NMC-333 cells (Fig. 5f), the cell with baseline electrolyte fails quickly within five cycles while the advanced electrolyte-based cell lasted for more than 50 cycles with a capacity retention of 59.2%, higher than any state-of-the-art reports (Table S2). This result further proves that the glassy Li metal is beneficial to achieve the best electrochemical performance for high-energy rechargeable Li batteries.

**Discussion and conclusion**

A combination of the cryogenic microscopic observations and molecular-level simulations presents an explicit picture of intermediate structural and morphological evolution of Li deposits from the atomic scale to a microscale particle for the first time. Depending on the atomic interaction during the initial nucleation (e.g. packing density, mass and energy transfer), the nanostructure of Li nuclei can vary from disorder to order, which eventually shape the final microstructure and affect the performance. To further illustrate this aspect, Fig. 6 provides an interesting simulation how deposition rate affects the morphology evolutions and kinetic variations in shaping the surface landscape and defect formation in the bulk. At a higher rate (Fig. 6a), the less incubation time could afford a more stabilized growth of c-Li in the EDLi to create larger grains in the microstructure. However, the mismatch of the orientation among the grains could result in rougher surface and a significant amount of defect formation from dislocations, grain boundaries to voids. As the Li deposition rate decreases, smoother surface and less defects in the bulk developed in the microstructure and morphology.

For rechargeable Li batteries, glassy Li has been proved to possess the desired nanostructure since it facilitates the formation of large Li grains and achieve high Coulombic efficiency. The absence of the order nanostructure and grain boundaries enables Li metal to grow along multi-dimensions rather than the epitaxial manner, and maintain good structural connection and reversibility during plating and stripping. Tuning the temporal and spatial confinements in mass and energy transfer by different strategies could help to obtain glassy Li metal deposits, including methods to lower current density, design advanced electrolyte composition and use 3D current collector. These strategies are able to alter the bulk microstructure of Li metal electrode and obtain more homogeneous large Li deposits with improved cycling performance.

Amorphous metals or metallic glasses are attractive as a class of advanced functional materials for applications in different technological areas and the scientific explorations on the glass formation and glassy phenomena need more attention [35, 36, 37, 38]. Conventional metallic glasses are made by alloying more than two metals especially transition metals by fast quenching (e.g. $>10^6$ K s$^{-1}$) [35]. No one before us has succeeded in synthesizing metallic glasses with a single component and with very reactive alkali metals. This work demonstrates that electrochemical deposition is a powerful and effective method to obtain such amorphous metals, including Na, K, Mg and Zn (Fig. S20-24). More importantly, the properties of such amorphous metals, amounts, and particle sizes and distributions could be tuned by adjusting the current density and deposition time through optimization. These new amorphous active metals will definitely open new opportunities to various applications besides the metallic glasses and energy storage fields, including biomedicine, nanotechnology, and microelectromechanical systems.

**Acknowledgments:**

We would like to acknowledge the UC Irvine Materials Research Institute (IMRI) for the use of Cryo-Electron Microscopy Facility and Kratos XPS, funded in part by the National Science Foundation Major Research Instrumentation Program under Grant No. CHE-1338173. We thank Dr. Xiaoqing Pan, Dr. Toshihiro Aoki, Dr. Li Xing and Dr. Jian-Guo Zheng for their assistance with the microscopy. Work at the Molecular Foundry was supported by the Office of Science, Office of Basic Energy Sciences, of the U.S. Department of Energy under Contract No. DE-AC02-05CH11231.The SEM was performed in part at the San Diego Nanotechnology Infrastructure (SDNI), a member of the National Nanotechnology Coordinated Infrastructure, which is supported by the National Science Foundation (Grant ECCS-1542148). We also would like to thank Prof. Adri van Duin of the Pennsylvania State University for providing the ReaxFF parameters used in this work. We appreciate Dr. Wu Xu and Dr. Marco Olguin for discussing the results.

**Funding:**

This work is supported by the Assistant Secretary for Energy Efficiency and Renewable Energy, Office of Vehicle Technologies of the U.S. Department of Energy in the Advanced Battery Materials Research (BMR) Program (Battery500 Consortium). INL is operated by Battelle Energy Alliance under Contract No. DE-DE-AC07-05ID14517 for the U.S. Department of Energy.

**Author Contributions:**

XW, SM, and BL conceived the idea. XW designed the experiments and conducted the cryo-TEM. YL, BL, MZ, and AB helped with cell making, electrochemical performance evaluation, EELS and SEM. XW, YL and SM analyzed and discussed the experimental data. XW and YL made the schematic images. GP, ED and BL developed the concept on $r$-MD approach. GP conducted the simulation work. BL provided the major interpretation and discussion of the $r$-MD results in the paper. XR and JZ provided the advanced electrolyte and the anode-free electrochemical performance. PL, JX, JL and ED participated in discussing the results and commenting on the manuscript. The manuscript was mainly written and revised by XW, BL, GP and SM. All authors have given approval to the final version of the manuscript.

**Competing interests:**

Authors declare no competing interests.

**References and Notes:**

1.  Steiger J, Kramer D, Mönig R. Mechanisms of dendritic growth investigated by in situ light microscopy during electrodeposition and dissolution of lithium. *J Power Sources* 2014, **261:** 112-119.

2.  Wood KN, Kazyak E, Chadwick AF, Chen K-H, Zhang J-G, Thornton K, *et al.* Dendrites and Pits: Untangling the Complex Behavior of Lithium Metal Anodes through Operando Video Microscopy. *ACS Central Science* 2016, **2**(11)**:** 790-801.

3.      Cohen YS, Cohen Y, Aurbach D. Micromorphological Studies of Lithium Electrodes in Alkyl Carbonate Solutions Using in Situ Atomic Force Microscopy. *J Phys Chem B* 2000, **104**(51)**:** 12282-12291.

4.      Mogi R, Inaba M, Iriyama Y, Abe T, Ogumi Z. In Situ Atomic Force Microscopy Study on Lithium Deposition on Nickel Substrates at Elevated Temperatures. *J Electrochem Soc* 2002, **149**(4)**:** A385-A390.

5.      Orsini F, Du Pasquier A, Beaudoin B, Tarascon JM, Trentin M, Langenhuizen N*, et al.* In situ Scanning Electron Microscopy (SEM) observation of interfaces within plastic lithium batteries. *J Power Sources* 1998, **76**(1)**:** 19-29.

6.      Cheng J-H, Assegie AA, Huang C-J, Lin M-H, Tripathi AM, Wang C-C*, et al.* Visualization of Lithium Plating and Stripping via in Operando Transmission X-ray Microscopy. *J Phys Chem C* 2017, **121**(14)**:** 7761-7766.

7.      Kushima A, So KP, Su C, Bai P, Kuriyama N, Maebashi T*, et al.* Liquid cell transmission electron microscopy observation of lithium metal growth and dissolution: Root growth, dead lithium and lithium flotsams. *Nano Energy* 2017, **32:** 271-279.

8.      Li Y, Li Y, Pei A, Yan K, Sun Y, Wu C-L*, et al.* Atomic structure of sensitive battery materials and interfaces revealed by cryo–electron microscopy. *Science* 2017, **358**(6362)**:** 506-510.

9.      Wang X, Zhang M, Alvarado J, Wang S, Sina M, Lu B*, et al.* New Insights on the Structure of Electrochemically Deposited Lithium Metal and Its Solid Electrolyte Interphases via Cryogenic TEM. *Nano Lett* 2017, **17**(12)**:** 7606-7612.

10.     Dubrovskii V. Fundamentals of Nucleation Theory.  *Nucleation Theory and Growth of Nanostructures*. Springer Berlin Heidelberg: Berlin, Heidelberg, 2014, pp 1-73.

11.     Scharifker B, Mostany J. Electrochemical Nucleation and Growth.  *Encyclopedia of Electrochemistry*, 2007.

12.     Ely DR, Jana A, García RE. Phase field kinetics of lithium electrodeposits. *J Power Sources* 2014, **272:** 581-594.

13.     Ely DR, García RE. Heterogeneous Nucleation and Growth of Lithium Electrodeposits on Negative Electrodes. *J Electrochem Soc* 2013, **160**(4)**:** A662-A668.

14.     Akolkar R. Modeling dendrite growth during lithium electrodeposition at sub-ambient temperature. *J Power Sources* 2014, **246:** 84-89.

15. Akolkar R. Mathematical model of the dendritic growth during lithium electrodeposition. *J Power Sources* 2013, **232:** 23-28.

16. Chazalviel JN. Electrochemical aspects of the generation of ramified metallic electrodeposits. *Phys Rev A* 1990, **42**(12)**:** 7355-7367.

17. Wang X, Li Y, Meng YS. Cryogenic Electron Microscopy for Characterizing and Diagnosing Batteries. *Joule* 2018, **2**(11)**:** 2225-2234.

18. Wang F, Graetz J, Moreno MS, Ma C, Wu L, Volkov V, *et al.* Chemical Distribution and Bonding of Lithium in Intercalated Graphite: Identification with Optimized Electron Energy Loss Spectroscopy. *ACS Nano* 2011, **5**(2)**:** 1190-1197.

19. Loh ND, Sen S, Bosman M, Tan SF, Zhong J, Nijhuis CA*, et al.* Multistep nucleation of nanocrystals in aqueous solution. *Nature Chemistry* 2017, **9**(1)**:** 77-82.

20. Yang J, Koo J, Kim S, Jeon S, Choi BK, Kwon S*, et al.* Amorphous-Phase-Mediated Crystallization of Ni Nanocrystals Revealed by High-Resolution Liquid-Phase Electron Microscopy. *J Am Chem Soc* 2019, **141**(2)**:** 763-768.

21. Liao H-G, Cui L, Whitelam S, Zheng H. Real-Time Imaging of Pt3Fe Nanorod Growth in Solution. *Science* 2012, **336**(6084)**:** 1011-1014.

22. Wittig JE, Bentley J, Allard LF. In situ investigation of ordering phase transformations in FePt magnetic nanoparticles. *Ultramicroscopy* 2017, **176:** 218-232.

23. Chi M, Wang C, Lei Y, Wang G, Li D, More KL*, et al.* Surface faceting and elemental diffusion behaviour at atomic scale for alloy nanoparticles during in situ annealing. *Nat Commun* 2015, **6**(1)**:** 8925.

24. van Duin ACT, Dasgupta S, Lorant F, Goddard WA. ReaxFF:  A Reactive Force Field for Hydrocarbons. *J Phys Chem A* 2001, **105**(41)**:** 9396-9409.

25. Plimpton S, Crozier P, Thompson A. LAMMPS-large-scale atomic/molecular massively parallel simulator. *Sandia National Laboratories* 2007, **18:** 43.

26. Majure DL, Haskins RW, Lee NJ, Ebeling RM, Maier RS, Marsh CP*, et al.* Large-Scale Atomic/Molecular Massively Parallel Simulator (LAMMPS) Simulations of the Effects of Chirality and Diameter on the Pullout Force in a Carbon Nanotube Bundle.  2008 DoD HPCMP Users Group Conference; 2008 14-17 July 2008; 2008. p. 201-207.

27. Stukowski A. Visualization and analysis of atomistic simulation data with OVITO–the Open Visualization Tool. *Modelling and Simulation in Materials Science and Engineering* 2009, **18**(1)**:** 015012.

28. Onofrio N, Strachan A. Voltage equilibration for reactive atomistic simulations of electrochemical processes. *The Journal of Chemical Physics* 2015, **143**(5)**:** 054109.

29. Jensen BD, Wise KE, Odegard GM. The effect of time step, thermostat, and strain rate on ReaxFF simulations of mechanical failure in diamond, graphene, and carbon nanotube. *J Comput Chem* 2015, **36**(21)**:** 1587-1596.

30. Cui Z, Gao F, Cui Z, Qu J. Developing a second nearest-neighbor modified embedded atom method interatomic potential for lithium. *Modelling and Simulation in Materials Science and Engineering* 2011, **20**(1)**:** 015014.

31. Stukowski A. Structure identification methods for atomistic simulations of crystalline materials. *Modelling and Simulation in Materials Science and Engineering* 2012, **20**(4)**:** 045021.

32. Baldan A. Review Progress in Ostwald ripening theories and their applications to nickel-base superalloys Part I: Ostwald ripening theories. *J Mater Sci* 2002, **37**(11)**:** 2171-2202.

33. Nanev CN. Theory of nucleation. *Handbook of Crystal Growth (Second Edition)*. Elsevier, 2015, pp 315-358.

34. Karthika S, Radhakrishnan TK, Kalaichelvi P. A Review of Classical and Nonclassical Nucleation Theories. *Cryst Growth Des* 2016, **16**(11)**:** 6663-6681.

35. Greer AL. Metallic Glasses. *Science* 1995, **267**(5206)**:** 1947-1953.

36. Jafary-Zadeh M, Praveen Kumar G, Branicio P, Seifi M, Lewandowski J, Cui F. A Critical Review on Metallic Glasses as Structural Materials for Cardiovascular Stent Applications. *Journal of Functional Biomaterials* 2018, **9**(1)**:** 19.

37. Chen M. A brief overview of bulk metallic glasses. *Npg Asia Materials* 2011, **3:** 82.

38. Khan MM, Nemati A, Rahman ZU, Shah UH, Asgar H, Haider W. Recent Advancements in Bulk Metallic Glasses and Their Applications: A Review. *Crit Rev Solid State Mater Sci* 2018, **43**(3)**:** 233-268.

**Data and materials availability:**

All data is available in the main text or the supplementary materials.

**Supplementary Materials:**

Materials and Methods

Figures S1-S24

Tables S1-S2

Movies S1-S6

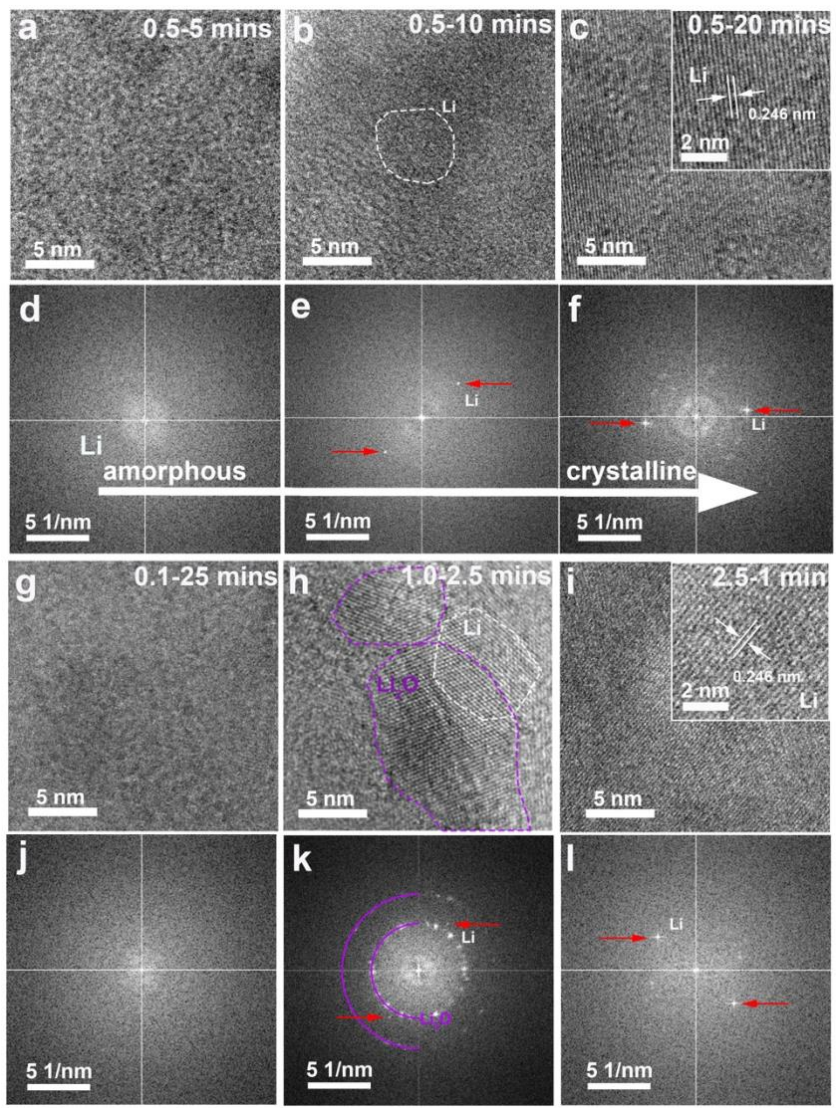

**Fig. 1.** Nanostructure evolution of Li deposit as a function of deposition time and current density. TEM images (a-c and g-i) and their corresponding FFT patterns (d-f and j-l) of the Li deposits at 0.5 mA cm$_{-2}$ for 5 min (a and d), 10 min (b and e) and 20 min (c and f), 0.1 ma cm$_{-2}$ for 25 min (g and j), 1.0 mA cm$_{-2}$ for 2.5 min (h and k) and 2.5 mA cm$_{-2}$ for 1.0 min (i and l). The images with Li metal lattice were obtained along the [001] zone axis and the characteristic bright diffraction spots of Li metal were highlighted by the red arrows.

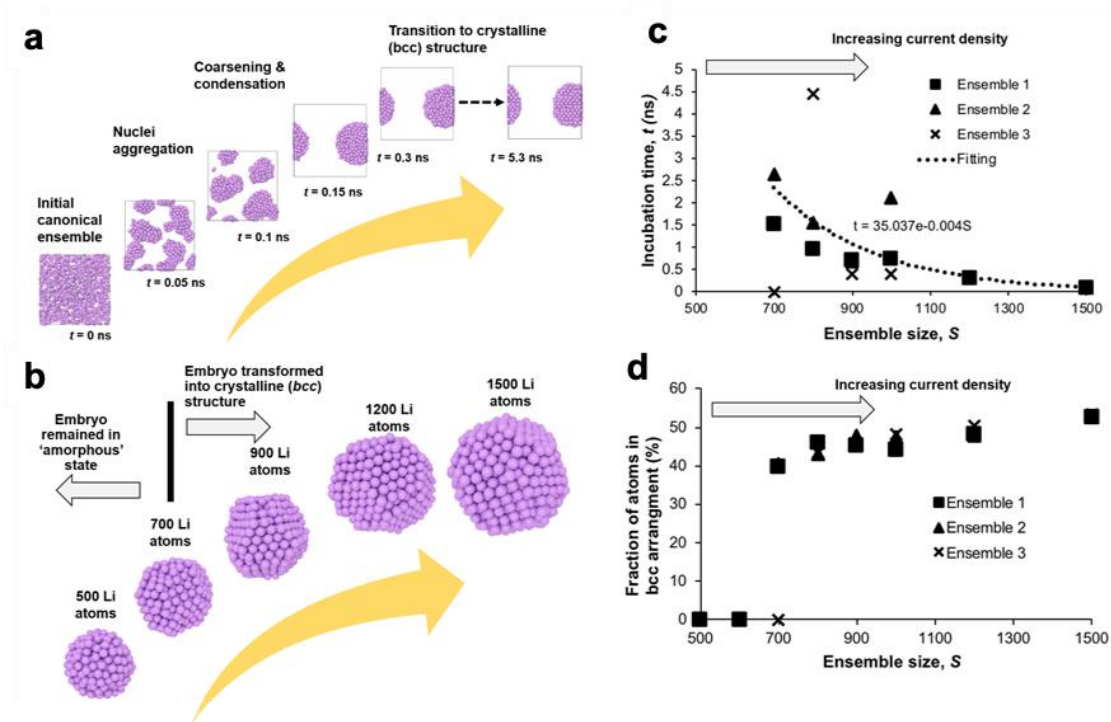

**Fig. 2.** MD simulation of the nucleation process of Li metal. (a) An example of the Li nucleation, growth and kinetic pathway to an embryo with an ensemble of 700 Li atoms. (b) A sampling of the final state of various sizes of embryos at the end of the simulation (5.3 ns). (c) Incubation time to second order phase transition as a function of ensemble size. (d) A fraction of Li atoms in bcc lattice arrangement as a function of ensemble size.

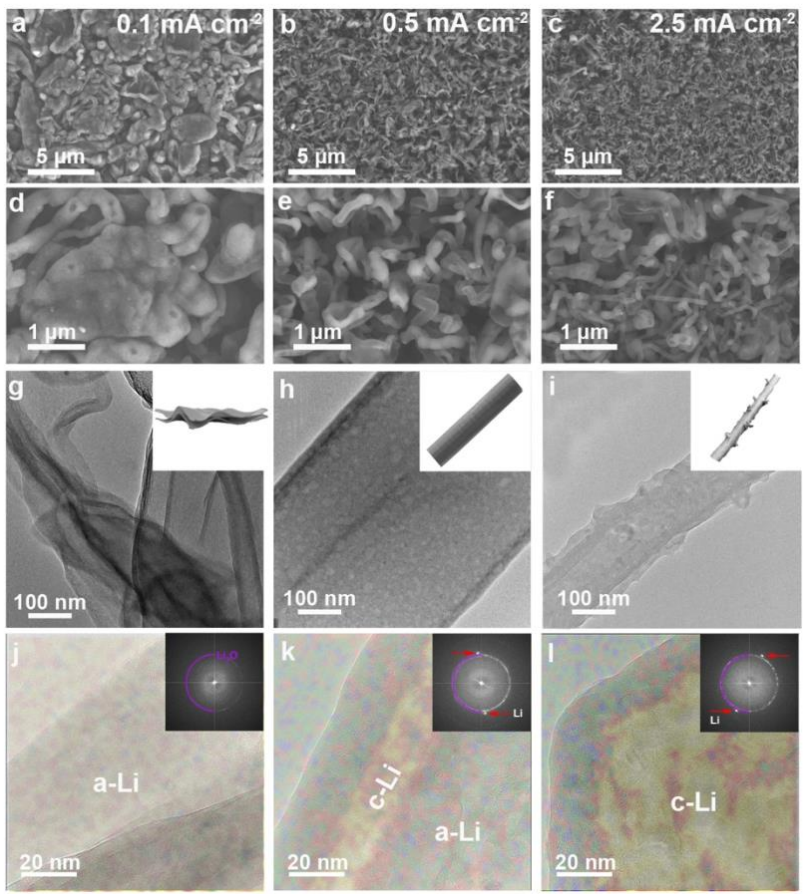

**Fig. 3.** Microstructure and morphology of Li deposits after 1 mAh cm$_{-2}$ plating. SEM (a-f) and cryo-TEM images (g-l) of the Li deposits at 0.1 mA cm$_{-2}$ for 10 h (a, d, g and j), 0.5 mA cm$_{-2}$ for 2 h (b, e, h and k), and 2.5 mA cm$_{-2}$ for 0.4 h (c, f, i and l). The images with Li metal lattice were obtained along the [001] zone axis and the characteristic bright diffraction spots of Li metal were highlighted by the white circles. Based on the characteristic diffraction spots of Li metal (highlighted by the red arrows), inverse FFT was carried out to show the distribution of crystalline Li in the corresponding images. The colorful distribution of crystalline Li was overlapped with the corresponding TEM image in j-l. The pristine images without color can be found in Fig. S18.

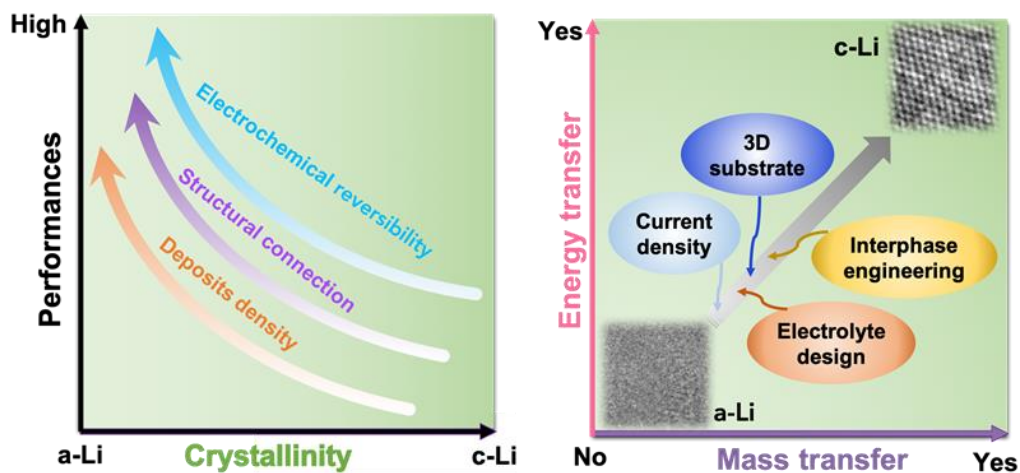

**Fig. 4.** Correlation between crystallinity of Li metal and performance (left) and strategies (right) to achieve better performance. The performance (left) is specified to the electrochemical performance of Li metal as an anode for Li-metal batteries, including the high Coulombic efficiency (CE), long cycling life, low volume change, and absence of Li dendrites. The structural connection is referred to the capability to maintain the electronic and ionic pathway for charge transfer and ion transportation; poor structural connection will facilitate to loss electrochemical activity and form 'dead' Li. The electrochemical reversibility is measured by the content ratio of the stripped Li by plated Li, which should be close to 100%. The ideal deposits density should be consistent with the theoretical density of Li metal (0.534 g cm$_{-3}$).

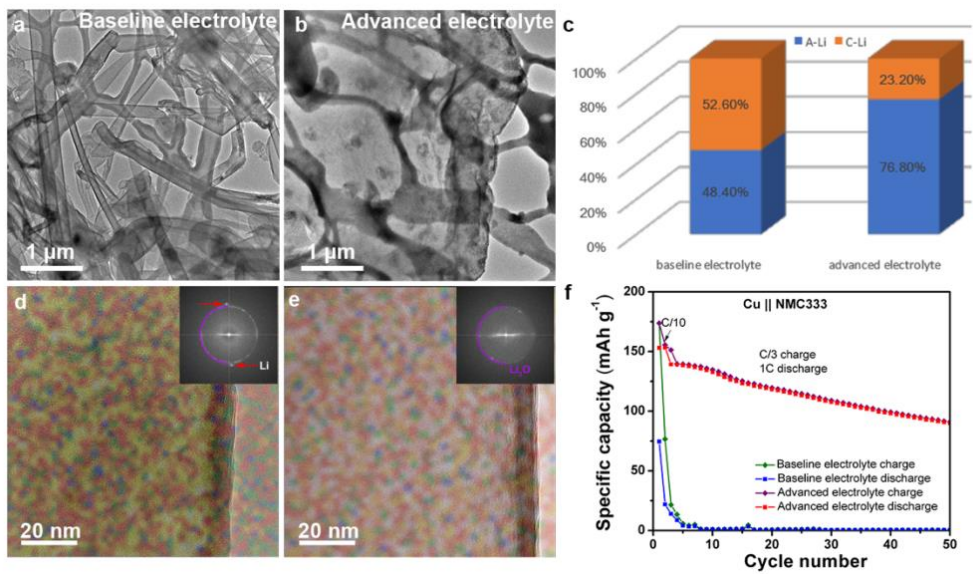

**Fig. 5.** The influence of electrolyte on the nano and micro structure of Li deposits and the resultant performance. Nano (d and e) and microstructure (a and b) of Li deposits from baseline electrolyte (a and d) and advanced electrolyte (b and e); statistic distribution (c) of amorphous Li (a-Li) and crystalline Li (c-Li) in two electrolytes; the cycling performance comparison (f) of two electrolytes by using Li-free anode Cu∥NMC-333 cells. The images with Li metal lattice were obtained along the [001] zone axis and the characteristic bright diffraction spots of Li metal were highlighted by the white circles. Based on the characteristic diffraction spots of Li metal (highlighted by the red arrows), inverse FFT was carried out to show the distribution of crystalline Li in the corresponding images. The colorful distribution of crystalline Li was overlapped with the corresponding TEM image in d and e. The pristine images without color can be found in Fig. S19. The statistical analysis was performed based on the ~100 separated random Li ribbons; c-Li was counted if there present lattice or FFT diffraction spots belonged to Li metal; otherwise a-Li was counted.

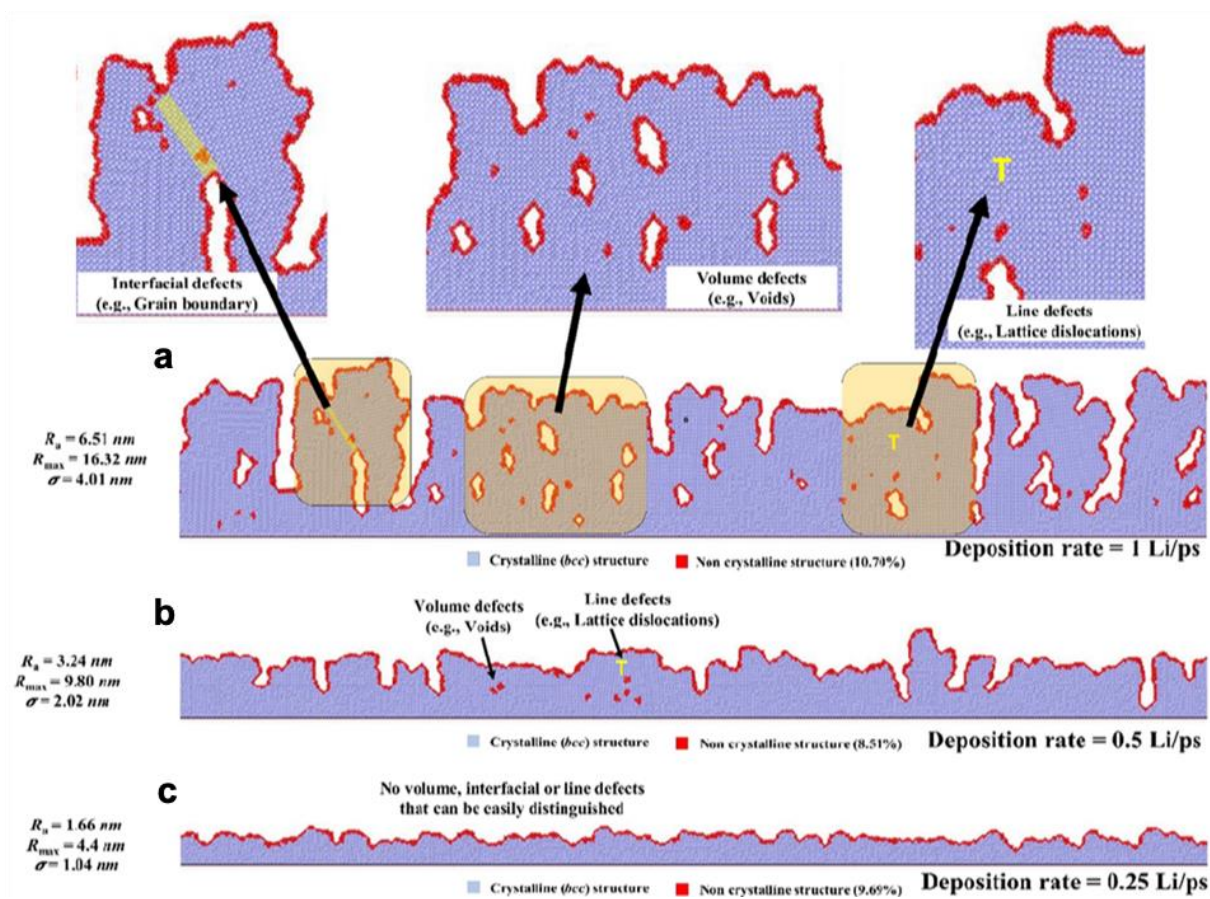

**Fig. 6.** Simulated results of Li deposition at three different rates: 1 Li ps−1, 0.5 Li ps−1, and 0.25 Li ps−1. A quasi-three-dimensional simulation domain of 112.3 (L) × 0.7 (D) × 35.1 (H) nm3 was used to simulate the rate effects on Li deposit. The MEAM force field was used to capture the Li atomic interactions. The LAMMPS simulation package was used to perform all atomistic simulations where the total simulation time was 45 ns with a time step of 1 fs. The quality and characteristics of the Li deposition are quantified by three parameters: an average surface roughness ($R_a$), the maximum surface roughness ($R_{max}$), and the standard deviation of surface roughness ($\sigma$).